\documentclass[reprint,aps,prl,twocolumn,superscriptaddress,longbibliography]{revtex4-1}

\usepackage{amsmath}
\usepackage{amssymb}
\usepackage{graphicx}
\usepackage{afterpage}
\usepackage{amsfonts}
\usepackage{amssymb}
\usepackage{graphicx}
\usepackage{braket}
\usepackage{textcomp}

\begin{document}

\title{Picosecond control of quantum dot laser emission by coherent phonons}

\author{T. Czerniuk}
\affiliation{Experimentelle Physik 2, Technische Universit\"at Dortmund, 44221 Dortmund, Germany}
\email{thomas.czerniuk@tu-dortmund.de}

\author{D.~Wigger}
\email{d.wigger@wwu.de}
\affiliation{Institut f\"{u}r Festk\"{o}rpertheorie, Universit\"{a}t M\"{u}nster, 48149 M\"{u}nster, Germany}

\author{A.~V.~Akimov}
\affiliation{School of Physics and Astronomy, University of Nottingham, Nottingham NG7 2RD, United Kingdom}

\author{C.~Schneider}
\author{M.~Kamp}
\author{S.~H\"{o}fling}
\affiliation{Technische Physik, Universit\"{a}t W\"{u}rzburg, 97074 W\"{u}rzburg, Germany}

\author{D. R. Yakovlev}
\affiliation{Experimentelle Physik 2, Technische Universit\"at Dortmund, 44221 Dortmund, Germany}
\affiliation{Ioffe Institute, Russian Academy of Sciences, 194021 St. Petersburg, Russia}

\author{T.~Kuhn}
\author{D.~E.~Reiter}
\affiliation{Institut f\"{u}r Festk\"{o}rpertheorie, Universit\"{a}t M\"{u}nster, 48149 M\"{u}nster, Germany}

\author{M.~Bayer}
\affiliation{Experimentelle Physik 2, Technische Universit\"at Dortmund, 44221 Dortmund, Germany}
\affiliation{Ioffe Institute, Russian Academy of Sciences, 194021 St. Petersburg, Russia}

\begin{abstract} 
A picosecond acoustic pulse can be used to control the lasing emission from semiconductor nanostructures by shifting their electronic transitions. When the active medium, here an ensemble of (In,Ga)As quantum dots, is shifted into or out of resonance with the cavity mode, a large enhancement or suppression of the lasing emission can dynamically be achieved. Most interesting, even in the case when gain medium and cavity mode are in resonance, we observe an enhancement of the lasing due to shaking by coherent phonons. In order to understand the interactions of the non-linearly coupled photon-exciton-phonon subsystems, we develop a semiclassical model and find an excellent agreement between theory and experiment. 
\end{abstract}

\maketitle

The enhanced light-matter-interaction of a semiconductor nanostructure, which is placed in an optical resonator, with the confined photonic field has paved the way to a large number of novel optical phenomena, both in the weak- \cite{PhysRevLett.81.1110,PhysRevLett.116.020401} and strong-coupling regime \cite{balili2007bos,PhysRevA.53.4250,schneider2013}. For the observation of any of these, the energy associated with the resonant photons needs to match the electronic transition of the gain material. Usually this needs to be arranged during fabrication of the structure, since there are only limited tools to achieve resonance post-growth. Recently, a new approach to dynamically shift the electronic transition of the gain material has been developed \cite{bruggemann2012las}, which is proposed to be useful for the study of a broad range of quantum phenomena \cite{Nomura2011}. This method is based on ultrafast mechanical vibrations: a broadband acoustic pulse containing coherent phonons up to THz frequencies passes through the gain medium and changes dynamically the transition energies, resulting in a strongly modified coupling to the optical resonator mode. The original experiment was performed on a microcavity laser with a quantum dot (QD) ensemble as the active medium, and has been extended to nanostructures like optically active quantum wires \cite{Shiri2012}, electronic transport devices \cite{PhysRevLett.108.226601}, and optomechanical resonators \cite{czerniuk2014las,LanzillottiKimura201580}.

To fully explore the potential of this method, we develop a theoretical model of the lasing dynamics in a microcavity laser system, which consists of three nonlinearly coupled subsystems: excitons, photons, and phonons. Experiments exploring several excitation regimes and detunings between QD ensemble and microcavity resonator accompany the theory, from which we find good agreement with simulations. Our combined approach allows us to understand the ongoing dynamics in detail. In particular, we show that we can distinguish between two effects: the first one is an adiabatic response of the lasing efficiency following the total number of QDs coupling to the resonator; the second one is a transient increase of the lasing output, when the initially off-resonant reservoir of excited QDs is shaken and guided into the cavity mode. These effects occur on different time scales, i.e., the adiabatic shift is efficient for phonons of any frequency, while the shaking effect requires frequencies comparable with the exciton lifetime in the lasing regime. This understanding is essential to enhance the technology and exploit ultrafast control of lasing using coherent phonons. 

\begin{figure}[ht]
\includegraphics{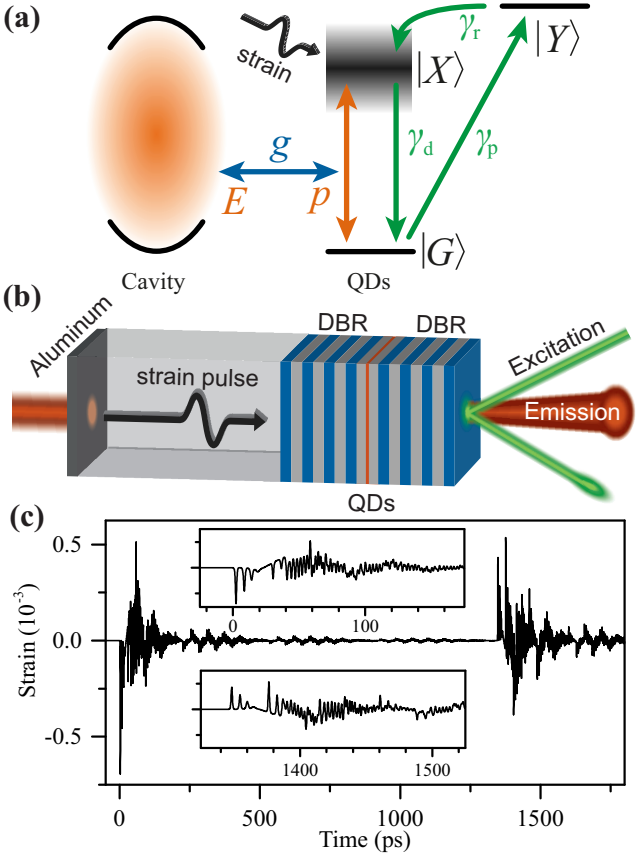}
\caption{(a) Sketch of the theoretical model. (b) Scheme of the experiment \cite{bruggemann2012las}. (c) Calculated strain profile $\eta(t)$ at the QD layer, showing the incident pulse at $t = 0$ and the reflected one at $t \approx 1.3$\,ns as close-ups in the insets.}
\label{fig:scheme}
\end{figure}

Sketches of the theory and the experiment are shown in Figs.~\ref{fig:scheme} (a) and \ref{fig:scheme} (b), respectively. Let us first focus on the lasing dynamics of the QD ensemble. We model each QD as a two-level system consisting of the ground state $\left|G\right>$ and the exciton state $\left|X\right>$. Due to the differences in size, the corresponding transition energies $\hbar\omega_i$ are modeled by a Gaussian centered at $\omega_{\rm{QD}}$ with a full width at half maximum (FWHM) of $\Delta_{\rm{QD}}$. We assume that the QDs are energetically closely spaced, such that we can use a continuous distribution $n(\omega)$. For each QD mode $\omega_i$ we employ a rate equation model to simulate the pump dynamics. For the pumping we include an additional energy level $|Y\rangle$, which can be thought of as a wetting layer state and which is populated from the ground state with a pump rate $\gamma_{\rm p}$ ($\left|G\right>\to\left|Y\right>$). The excitation then relaxes in the exciton state via the relaxation rate $\gamma_{\rm r}$ ($\left|Y\right>\to\left|X\right>$). From the exciton state spontaneous decay into the ground state occurs, which is described by the decay rate $\gamma_{\rm d}$ ($\left|X\right>\to\left|G\right>$). Each QD is coupled to the cavity mode $E$ with the frequency $\omega_{\rm c}$ via the coupling element $g$ in the usual dipole, rotating wave and slowly varying amplitude approximation.

Initially the electric field is given by white noise. When the QD is inverted, a polarization $p_{\rm {\omega}}$ between the ground state and the exciton builds up. The polarization is determined by the inversion and the detuning between each QD transition $\omega_i$ and the cavity mode $\omega_{\rm c}$, i.e., it is strongest, when the QD is resonant with the cavity $\omega_i=\omega_{\rm c}$. Note that the polarization dephases due to the pump and decay. Further, there is an additional polarization dephasing contribution \cite{chow2005the}, which is accounted for by the rate $\gamma$. Including the cavity loss by the rate $\gamma_{\rm l}$, the electric field dynamics is
\begin{equation}
\frac{{\rm d}E}{{\rm d}t} = -\gamma_{\rm l} E + i g \int n(\omega) p_{\rm {\omega}}(t)\, {\rm d}\omega {\rm,}
\end{equation}
where $E(t)$ and $p_{\rm {\omega}}(t)$ are in a frame rotating with the cavity frequency. Here, we see that the density, inversion and actual detuning of the QDs via the polarizations are important for the strength of the electric field.

In the experiment, the same laser like in Ref.~\cite{bruggemann2012las} is studied. The microcavity resonator is made of two distributed Bragg reflectors (DBRs) sandwiching a GaAs cavity layer with a variable thickness, where optically pumped (In,Ga)As QDs are placed. While the linewidth of the cavity mode is only 1.2\,meV, the broadening of the QD ensemble is 11\,meV, resulting in an inefficient coupling (see SOM). For the calculations we choose the parameters of the QD ensemble according to the experimental setting and simulate $N=5\times 10^{4}$ QDs. The cavity mode is set to the value from the experiment and its width is used to determine the cavity loss rate $\gamma_{\rm l}=0.4$~ps$^{-1}$. For the lasing dynamics, we take the parameters giving the best agreement with experiment $\gamma_{\rm d}=0.03$~ps$^{-1}$, $\gamma_{\rm r}=0.5$~ps$^{-1}$, $\gamma_{\rm}=1$~ps$^{-1}$ and a laser coupling constant of $g=2.8$~ps$^{-1}$, which are consistent with established theoretical models \cite{chow2005the,chow2014emi}. The lasing threshold is determined by $\gamma_{\rm d}$, such that we typically look at the pump rate in comparison to this with $\Gamma=\gamma_{\rm p}-\gamma_{\rm d}$. 

To modify the lasing properties, a coherent phonon pulse is impinged on the QD ensemble. In experiment, the phonons are generated by optical excitation of a 100\,nm aluminum film, which is deposited on the backside of the sample. Onto the aluminum film a short, high energetic laser pulse is focused. Due to rapid thermal expansion following the light absorption, a few picosecond long acoustic pulse of coherent phonons is launched and subsequently injected into the (100)-GaAs substrate \cite{thomsen1986sur}. To prevent a strong scattering of the coherent phonons, the sample is placed into a cryostat and cooled down to 8\,K. During the acoustic pulse's propagation through the 100\,\textmu m thick substrate, non-linear and dispersive crystal effects stretch the pulse and lead to the formation of phonons with frequencies of up to several hundred GHz \cite{PhysRevB.64.064302}. Figure \ref{fig:scheme} (c)  shows the evolution of the strain $\eta(t)$ at the QD layer, which was calculated using the transfer-matrix and scattering states method \cite{PhysRevB.60.15554,PhysRevB.38.1427}. Two pulses can be distinguished: the first one is the incident pulse coming from the substrate at $t=0$ and the second one is its reflection from the front surface of the sample. It passes the QD layer at $t\approx 1.3\,\rm{ns}$ according to twice the transit time through the top DBR. Note that the reflected strain pulse has flipped its sign at the open surface.

The strain field $\eta(t)$ changes the transition energy of every QD via $\hbar \omega_i \to \hbar \omega_i  + D \eta(t)$. In the simulations we take $D=-10$~eV as the deformation potential coupling constant \cite{PhysRevB.71.235329} and define the instantaneous detuning of the QD ensemble with respect to the cavity mode as 
\begin{equation}
	 \Delta(t) =\hbar\omega_{\rm c} - [\hbar \omega_{\rm QD}  + D \eta(t)]\, .
\end{equation}

When passing the QD layer, the induced energy shift results in a change in the emission intensity of the laser that is detected with a streak camera with a time resolution of $25$~ps. In the simulation we therefore integrate the electric field over a cosine-squared time window with a FWHM of $25$~ps.

Simulations and experiments were performed for three different detunings between the QD ensemble and the cavity mode: a large positive and a negative detuning and an almost resonant case. For each detuning, two different pump rates denoted by $P$ (experiment) and $\Gamma$ (simulation) are studied, which are expressed in terms of the respective lasing thresholds to provide comparable situations in terms of physics. We note that slightly different excitation powers relative to the threshold had to be used for best agreement. We assign this difference to different input-output curves measured and calculated (cf. the SOM), which will be discussed in more detail below. In the following, the upper panels (a) of each figure show the experimentally measured normalized emission intensity, while in the lower panels (b) the theoretical simulations are displayed. 
 
\begin{figure}[t]
\includegraphics{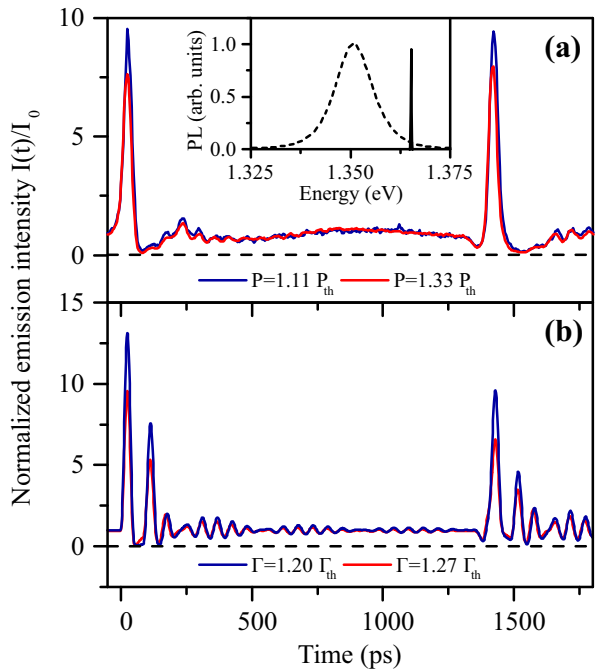}
\caption{(a) Measurement and (b) simulation of the dynamics of the lasing intensity under influence of the acoustic pulse for a detuning of 14.5\,meV. The inset shows the QD ensemble (dashed) and the cavity mode (solid) spectra.}
\label{fig:pos}
\end{figure}

First, we consider the case when the cavity mode is positively detuned from the QD ensemble by 14.5\,meV, similar to the case in Ref.~\cite{bruggemann2012las}. The dynamics of the lasing intensity are shown in Fig.~\ref{fig:pos} for two pump intensities slightly above the threshold. For both pump intensities, we see a strong amplification of the lasing in the experiment, when the incoming and reflected strain pulses hit the QD ensemble. This is well reproduced by the theory, which shows also two clear peaks at these times. The reason for the amplification is that negative parts (compression) of the strain pulse blueshift the QD ensemble, thereby shifting it towards the cavity. Also more details of the experiment can be reproduced by our model, e.g. after the amplification there is a quenching followed by smaller oscillations.

Taking a closer look, we see deviations of the simulated curve from the experimental data, e.g. the subsequent peaks following the two leading ones per pulse are rather distinct in theory but quite faint in experiment. These deviations may be explained by effects included in the model in a simplified way or even neglected due to the complexity of the underlying physics. These are treatment of carrier relaxation in the three-level model and neglecting possible multiexciton effects, acoustic wave damping and coupling to resonator modes such as the guided waves. These factors will lead to a broadening of the peaks blurring somewhat the measured signal and also explain the slightly larger enhancement in theory than in experiment. However, the good overall agreement underlines that the most important effects in the complex laser dynamics are captured by the theoretical treatment.

Another interesting aspect is that for higher pump intensity (red curve), the enhancement tends to become smaller, which is clearly reproduced by theory. In the highly nonlinear regime close to the lasing threshold, the system is very sensitive to coherent phonons and the control of the emission is most efficient.

\begin{figure}[t]
\includegraphics{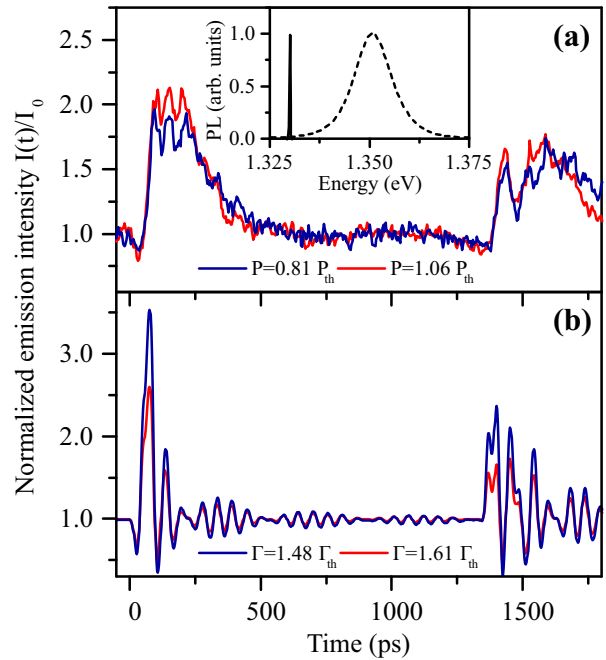}
\caption{Same as Fig.~\ref{fig:pos}, but for a detuning of $-17.8$\,meV.}
\label{fig:neg}
\end{figure}

With this in mind, we now look at a negatively detuned cavity mode with a detuning of $\Delta(0)=-17.8$~meV. We expect the lasing dynamics to be quite similar, since also here the phonons tune more QDs into resonance and thus enhance the lasing, now for positive strain. Indeed, we see that we have two large enhancements, one from the incoming strain pulse around $t=0$ and one at the reflected pulse around $t=1.3$~ns. However, there are differences in the response for this detuning. The incoming strain pulse [Fig.~\ref{fig:scheme}~(c)] starts with a strong negative part corresponding to a blueshift of the QDs. For the redshifted QDs discussed previously (Fig.~\ref{fig:pos}), this results in a strong enhancement of the lasing, while here the blueshifted QDs (Fig.~\ref{fig:neg}) are pushed even further away from the cavity. Accordingly, the lasing in the first case starts with an enhancement, while the lasing for this case starts with a quenching and only afterwards the output is enhanced. For the reflected pulse the sequence is inverted. Moreover, the total enhancement in the experiment with the redshifted QDs is about three times higher, due to the fact that the absolute detuning is smaller. In addition, the maximum of negative strain in the incoming pulse is slightly higher than its equivalent of positive strain in the reflected pulse. Thus, the negative part can compensate a larger detuning.

\begin{figure}[t]
\includegraphics{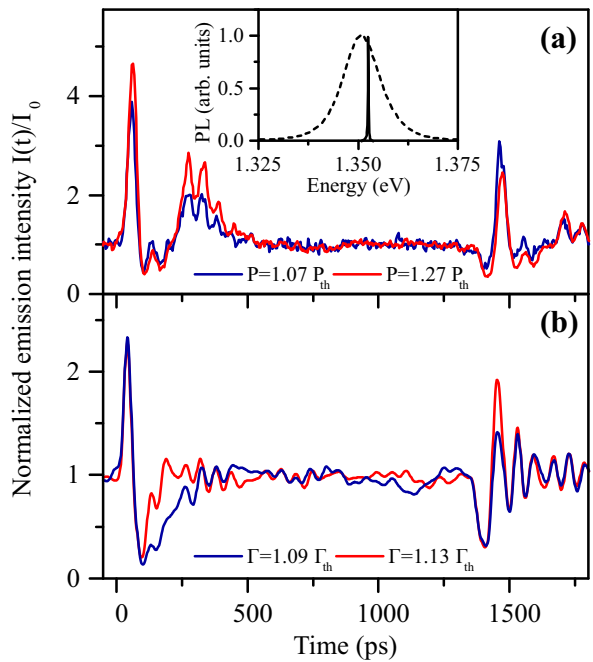}
\caption{Same as Fig.~\ref{fig:pos}, but for a detuning of 1.5\,meV in the experiment and 3.0\,meV in the simulations.}
\label{fig:res}
\end{figure}

The final measurement was taken for an almost resonant case of $\Delta(0)=1.5$~meV, for which $\Delta(0)=3$~meV is assumed in the theoretical curve in Fig.~\ref{fig:res} to achieve reasonable agreement. In particular the observed quenching requires this adjustment and it will be shown below that in the case of an even smaller detuning only intensity enhancements remain. When the strain pulse hits the QD ensemble, there is a strong enhancement, seen in both experiment and theory. Then a long period of quenching follows, while afterwards sizable oscillations are observed. Let us compare this to the profile of the strain pulse [cf.~Fig.~\ref{fig:scheme}~(c)]. The first soliton-like peaks of the strain pulse around $t=0$ shift the QD ensemble very rapidly into resonance followed by an oscillatory part, which on average increases the detuning. The second regime is given for the more regular oscillations in the strain pulse for times $200$~ps$~<t<1200$~ps, where the QDs follow the shift adiabatically. The lasing emission shown in Fig.~\ref{fig:res} reflects these oscillations. Also the asymmetry in the incoming and reflected pulse is observed in the simulations. 

\begin{figure}[t]
\includegraphics{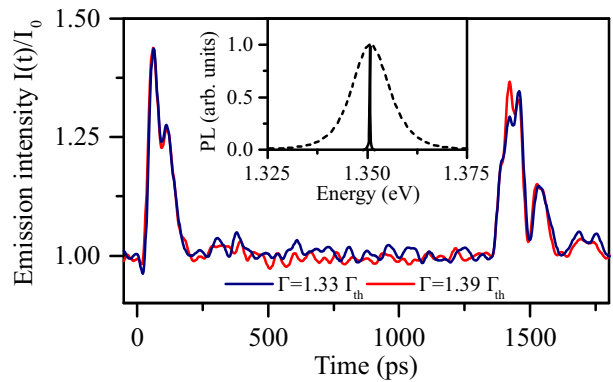}
\caption{Simulation of the laser emission for zero initial detuning.}
\label{fig:zero}
\end{figure}

An open question is still the relative contribution of the two fundamentally different mechanisms mentioned earlier - namely the adiabatic modulation, when coherent phonons of any frequency overcome an initial detuning, and the transient shaking effect. The latter is due to fast coherent phonons, which shift the QD transitions very rapidly, such that the spectral hole in the QD population due to the lasing is subsequently blue- and redshifted with respect to the cavity mode. In this way, highly excited formerly off-resonant QDs can contribute to the lasing and the spectral hole is artificially broadened.

In the experiment the two mechanisms cannot be distinguished: both contribute. To get some insight, we use our theoretical model for the special case of zero detuning, when the maximum of the QD distribution is already in resonance with the cavity mode. Here, we would expect that the strain pulse can only detune the QD ensemble, thus, the adiabatic contribution leads to quenching only. However, in the simulations shown Fig.~\ref{fig:zero} we see that for any pump rate above the lasing threshold, a significant enhancement of the lasing emission is obtained at $t\approx 0.05$~ns and at $t\approx 1.4$~ns corresponding to the times, when the fast oscillatory part passes the QD layer. Here, the dominant shaking effect does clearly overcome the adiabatic response and we conclude that the shaking effect is important to describe the laser dynamics.

Besides the detuning, another crucial input parameter is the pump intensity. In experiment, the lasing threshold is defined as the first kink, where the output exceeds the spontaneous emission. This threshold region is quite extended until full lasing is reached \cite{bruggemann2012las}. In contrast, in our semiclassical model there is a steep set-in of lasing at the threshold and already small variations of the pump intensity close to the threshold modify the lasing response significantly (see SOM). To include a broader threshold region, one needs to go to a fully quantum mechanical model to account for spontaneous emission\cite{chow2014emi}, which is extremely challenging when also including phonons. Moreover there are lasing parameters like the relaxation rate $\gamma_{\rm{r}}$, which are not easily experimentally accessible and have a significant impact on the response in our non-nonlinear model.

In conclusions, we have shown theoretically and experimentally that strain can be used to control the light-matter interaction on an ultrafast time scale in a QD microcavity laser. For a QD ensemble initially detuned with respect to the cavity mode, we find a strong amplification of the emission intensity. Even when the cavity mode is resonant on the QD ensemble, an amplification is found, underlining the effect of shaking on the QDs. It is appealing to work in the threshold region, where the shaking of QDs has the largest impact on the emission intensity. Our model allows us to study specifically tailored strain pulses to fully explore control of light-matter interaction by coherent phonons. 

This work was supported by the Deutsche Forschungsgemeinschaft (TRR 142) and the state of Bavaria. A.V.A. acknowledges the Alexander von Humboldt Foundation. M.B. acknowledges partial financial support from the Russian Ministry of Science and Education (contract No.14.Z50.31.0021). D.E.R.  acknowledges partial financial support from the German Academic Exchange Service (DAAD) within the P.R.I.M.E. program.


%

\end{document}